\documentclass[aps,prc,twocolumn,groupedaddress]{revtex4-1}

\usepackage{graphicx}  
\usepackage{epstopdf}
\usepackage{subfigure}
\usepackage{multirow}

\usepackage{fancyhdr}
\usepackage{longtable}
\usepackage{parskip}
\usepackage[T1]{fontenc}
\usepackage{dcolumn}   

\usepackage{bm}        
\usepackage{amsfonts}  
\usepackage{amsmath}   
\usepackage{amssymb}   
\usepackage{siunitx}
\usepackage{tabularx}
\usepackage{setspace}
\usepackage{indentfirst}

\newcommand{\pwisein}{\left\{ \begin{array}{ll}}
\newcommand{\pwiseout}{\end{array}\right.}

\newcommand{\bb}[1]{\raisebox{-3ex}[0pt][0pt]{\shortstack{#1}}}

\begin{document}

\title{ Thermonuclear fusion and generalized Gamow penetrability factor in intense laser fields }

\author{Jintao Qi}

\email[\textsuperscript{}]{qijintao18@gscaep.ac.cn}

\affiliation {Graduate School, China Academy of Engineering Physics, Beijing 100193, China}

\date{today}

\begin{abstract}
A theoretical study on thermonuclear fusion including deuterium-tritium (D-T) fusion and D-$^{3}$He fusion in intense laser fields has been shown in this article. With the laser fields expected to be available in the near future, some quantitative results for the laser-induced modifications to the cross-sections are given. 
It is reported that the cross-sections are more sensitive to the external laser fields at the lower energies.
An explicit generalized form of the Gamow penetrability factor is given for the predictions of the laser-induced effects for some other similar nuclear processes.
\end{abstract}

\maketitle

\section{Introduction}

Thermonuclear fusion is the conversion of two light nuclei to a heavier nucleus in a thermal environment, with the consequent release of a relatively large amount of energy. 
Thermonuclear fusion is a typical quantum tunneling process happening at extremely high temperatures typically about 10 keV (1 keV $\approx$ 1.16$\times$10$^{7}$ K). 
When fusion happens, the high temperatures give the two nuclei enough energy to overcome the Coulomb barrier.  
Similar quantum tunneling processes have been studied in some other areas of physics, such as tunneling ionization of atoms \cite{Ammosov1986,Corkum1996,Chen2000} and cluster radioactivity of heavy nuclei \cite{Delion2006,Poenaru1985,Saidi2015}. But the potential barriers felt by nuclei are much higher than those felt by electrons.

The invention of the chirped pulse amplification technique \cite{Strickland1985} has greatly promoted the development of the laser technologies. Intense laser fields at a peak intensity exceeding $10^{23}$ W/cm$^{2}$ may be generated soon with the extreme light infrastructure (ELI) in Europe \cite{Ur2015,Bala2017,Bala2019} or the superintense ultrafast laser facility (SULF) of Shanghai \cite{Li2018, Yu2018, Zhang2020}. The possible available intense laser fields have attracted some attention on the laser-induced effects on nuclear processes including $\alpha$ decay \cite{Keitel2013,Misicu2013,Delion2017,Qi2019,Palffy2019,Qi2020} and D-T fusion \cite{Queisser2019,Lv2019,Wang2020}.
For D-T fusion in the limit of very high laser frequencies as in x-ray regime, some qualitative estimations have been given by Queisser and Sch\"utzhold based on a Floquet scattering method \cite{Queisser2019} and by Lv $et$ $al.$ using a Krammer-Henneberger approximation \cite{Lv2019}. Yet these methods or approximations are not available in the near-infrared regime. 
In Ref. \cite{Wang2020}, Wang introduced the Volkov state to study the energy transferring effects of the D-T system in intense laser fields. In laser fields with intensities on the order below 10$^{21}$ W/cm$^{2}$, it was emphasized that the cross-section function does not change and the cross-sections are enhanced only due to the laser-induced energy-changes before tunneling. It was also reported that intense low-frequency laser fields in the near-infrared regime are highly effective in transferring energy to the D-T system.

For the lasers in the near-infrared regime from the majority of intense laser facilities such as ELI and SULF, the electric field strength corresponding to 10$^{24}$ W/cm$^{2}$ is comparable to the Coulomb field strength from a nucleus at a distance of about 100 fm. It is quite possible that the intense laser
fields can influence the Coulomb tail of nuclear potential. Indeed, recent works have considered the possible influences of intense laser fields on nuclear fission processes for heavy nulcei\cite{Qi2019,Palffy2019,Qi2020}. Notably, thermonuclear fusion may be more likely to be influenced by external intense laser fields, for the reason that the Coulomb barrier of light nuclei fusion is much lower than that of heavy nuclei fission. 

The goal of the current article is to study the possible influence of intense laser fields on thermonuclear fusion processes including D-T fusion and D-$^{3}$He fusion. In this article, the questions that how the external laser fields influence the thermonuclear fusion processes and how to calculate the laser-induced modification to the cross-section quantitatively are answered. 
The numerical results show that the laser-induced modifications to the cross-sections can reach about 79.1$\%$ for D-T fusion and 731.7$\%$ for D-$^{3}$He fusion, at the temperature of 3 keV. And at the temperature of 10 keV, the modifications to the cross-sections for D-T fusion and D-$^{3}$He fusion are 3.2$\%$ and 13.4$\%$, respectively. An explicit generalized form of the Gamow penetrability factor is given for the predictions of the laser-induced effects for some other similar nuclear processes.

This article is arranged as follows. In Sec. \uppercase\expandafter{\romannumeral2}, the numerical method will be presented. The subsections that the relative motion in external laser fields, the approximation method, penetrability and the cross-section in intense laser fields will be included. Numerical results, analyses, and discussions will be shown in Sec. \uppercase\expandafter{\romannumeral3}. A conclusion will be given in Sec. \uppercase\expandafter{\romannumeral4}.

\section{Method}

\subsection{Relative motion in intense laser fields}

The two-nuclei fusion system in an intense laser field can be described by the time-dependent Schr\"odinger equation (TDSE)
\begin{equation}\label{eq.s1}
i\hbar\frac{\partial\psi\left(\vec{r}_{1},\vec{r}_{2},t\right)}{\partial t}=H\left(t\right)\psi\left(\vec{r}_{1},\vec{r}_{2},t\right).
\end{equation}
The hamiltonian can be expressed as
\begin{equation}\label{eq.h1}
\begin{split}
H(t)= \sum\limits_{i=1,2} \frac{1}{2m_{i}}\left[\vec{p}_{i}-q_{i}\vec{A}(t)\right]^{2}+V(r),
\end{split}
\end{equation}
where $r=\left|\vec{r}_{1}-\vec{r}_{2}\right|$ denotes the distance between the two nuclei.

The size of a typical nucleus is on the order of 1 fm, which is much smaller than the wavelength of the near-infrared intense laser fields. Therefore a radiation gauge has been used so that the spatial dependency of the vector potential can be neglected reasonable.

For simplicity, the TDSE should be transformed into the center-of-mass (CM) coordinates as presented in \cite{Qi2020}. It is also convenient to study the two-nuclei fusion processes in intense laser fields with the TDSE for the relative motion 

\begin{equation}\label{23}
	i\hbar\frac{\partial\phi(\vec{r},t)}{\partial t}=\left[ -\frac{\hbar^2}{2\mu}\nabla_{r}^{2}+V(r)-q_{eff}\vec{r}\cdot\vec{\varepsilon}(t) \right]\phi(\vec{r},t),
\end{equation}
where $\mu=m_{1}m_{2}/(m_{1}+m_{2})$ denotes the reduced mass, and $q_{eff} = (q_1m_2-q_2m_1)/(m_1+m_2)$ is an effective charge for relative motion.

\begin{figure}[t]
	\centering
	\includegraphics[scale=0.33]{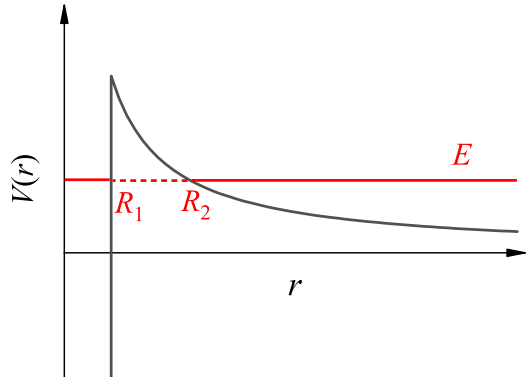}
	\caption{An illustration potential between two nuclei. The nuclear part of the potential used here is a spherical square well with radius $R_{1}$. The red line is the corresponding incident energy $E$. The tunneling entrance point $R_{2}$ can be obtained from the condition $V(R_{2})=E$.} \label{f.potential}
\end{figure}

The potential between the two nuclei includes the short-range of nuclear force and the long-range of Coulomb force which can be described as
\begin{equation}\label{eq.v0}
V(r)= \begin{cases} 
\begin{split}
&\frac{q_{1}q_{2}}{r},&r\geq R_{1};\\
&-U_{0},&r<R_{1},
\end{split}
\end{cases}
\end{equation}
where $R_{1}=1.13(a_{1}^{1/3}+a_{2}^{1/3})$ is the geometrical touch distance between the two nuclei, and $a_{1}$ and $a_{2}$ are the mass numbers of the two nuclei. Additionally, the value $U_{0}$ is usually $30-50$ MeV.

The interaction potential between the laser field and the relative motion nucleus is
\begin{align} \label{eq.vi}
		V_{I}(\vec{r},t) = -q_{eff}\vec{r}\cdot\vec{\varepsilon(t)} = -q_{eff}r\varepsilon(t)\cos\theta.
\end{align}
Note that the laser field is assumed to be linearly polarized along the $z$ axis and $\theta$ is the angular between the relative motion direction and the $+z$ axis.

\subsection{Quasistatic approximation}

Tunneling ionization of atoms in intense laser fields has been extensively studied \cite{Ammosov1986,Corkum1996,Chen2000}. The so-called tunneling domain is usually described by the limit that $\gamma \ll 1$. Here $\gamma = \sqrt{2mI_{p}}\omega/(q\varepsilon_{0})$ is the famous Keldysh parameter \cite{Keldysh1965} and often interpreted as $\gamma = \omega/\omega_{t}$ which is a radio of the frequency of laser $\omega$ to the frequency $\omega_{t}$ of electron tunneling through a potential barrier. The tunneling limit $\gamma \ll 1$ implies that when electron tunnels through a potential barrier, the laser field does not have time to change appreciably. Hence the tunneling limit is also mentioned as the quasistatic approximation which was discussed previously by us in Ref. \cite{Qi2019}. 

For the cases of thermonuclear reactions in the near-infrared intense laser fields, the quasistatic approximation is also applicable. For a thermal environment with a temperature of 10$^{8}$ K, the corresponding energy of the incident nucleus is about 10 keV. For an 800 nm near-infrared laser field adopted in this article, the corresponding photon energy is 1.55 eV. It can be calculated that the Keldysh parameters for D-T fusion and D-$^{3}$He fusion $\gamma=\sqrt{2\mu E}\omega/(q_{eff}\varepsilon_{0})=0.068$. 
Therefore the thermonuclear reactions in the near-infrared intense laser fields in this article can be discussed in the quasistatic regime.

\subsection{Penetrability and cross-section in intense laser fields}
A very important quantity to analyze nuclear reactions is the cross-section which describes the probability that a nuclear reaction will occur.
The fusion cross-section can be defined as
\begin{equation}
	\sigma(E)=\frac{S(E)}{E}\exp\left(-\frac{B_{G}}{\sqrt{E}}\right),
\end{equation}
where $E$ is the energy in the CM frame and the term $1/E$ is called the geometrical factor. $S(E)$ is called the astrophysical function (S-function) which relates the nuclear part of fusion reaction and varies slowly with energy \cite{Burbidge1957}. A 9-parameters S-function proposed by Bosh and Hale is used here \cite{Bosh1992}
\begin{equation}
	S(E)=\frac{A_{1}+E \left\lbrace A_{2}+E[A_{3}+E(A_{4}+EA_{5})] \right\rbrace}{1+E\left\lbrace B_{1}+E[B_{2}+E(B_{3}+EB_{4})] \right\rbrace},
\end{equation}
where the parameters $A_{i}$'s and $B_{i}$'s can be obtained in TABLE \uppercase\expandafter{\romannumeral4} of Ref. \cite{Bosh1992}. 
The exponential term $\exp(-B_{G}/\sqrt{E})$, the so-called Gamow penetrability factor, describes the probability of tunneling through the potential barrier without laser fields. The Gamow penetrability factor is an approximate formula based on the Wentzel-Kramers-Brillouin (WKB) method. 

From the quasistatic picture, it is still convenient to calculate the penetrability in an intense laser field for each electric strength $\varepsilon$ using the WKB method as

\begin{align} \label{eq.penetrability}
& P(E,\varepsilon,\theta) \nonumber \\
& = \exp\left\lbrace -\frac{2\sqrt{2\mu}}{\hbar}\int_{R_{1}}^{R_{2}}\sqrt{ V(r)-E+V_{I}(r,\varepsilon,\theta) }dr \right\rbrace,
\end{align}
where $V(r)$ and $V_{I}(r,\varepsilon,\theta)$ have been given in Eq. (\ref{eq.v0}) and Eq. (\ref{eq.vi}).

Hence, the fusion cross-section in an intense laser field can be written as 
\begin{equation} \label{eq.cross-section}
\sigma(E,\varepsilon,\theta)=\frac{S(E)}{E}P(E,\varepsilon,\theta).
\end{equation}

To analyze the laser-induced modification to fusion cross-section, a relative change of the cross-section is defined as
 \begin{equation} \label{eq.delta}
 	\Delta_{\sigma}(E,\varepsilon,\theta)=\frac{\sigma(E,\varepsilon,\theta)-\sigma(E,\varepsilon=0)}{\sigma(E,\varepsilon=0)}.
 \end{equation}
Note that the cross-section at each $\theta$ is assumed to be the same when there is no external laser field, so that the $\theta$ is neglected when $\varepsilon=0$.

\section{Results and discussions}

\subsection{Angular differential modifications to the cross-section}

\begin{figure}[htbp] 
	\centering
	\begin{minipage}[b]{0.23\textwidth}
		\includegraphics[width=1\textwidth]{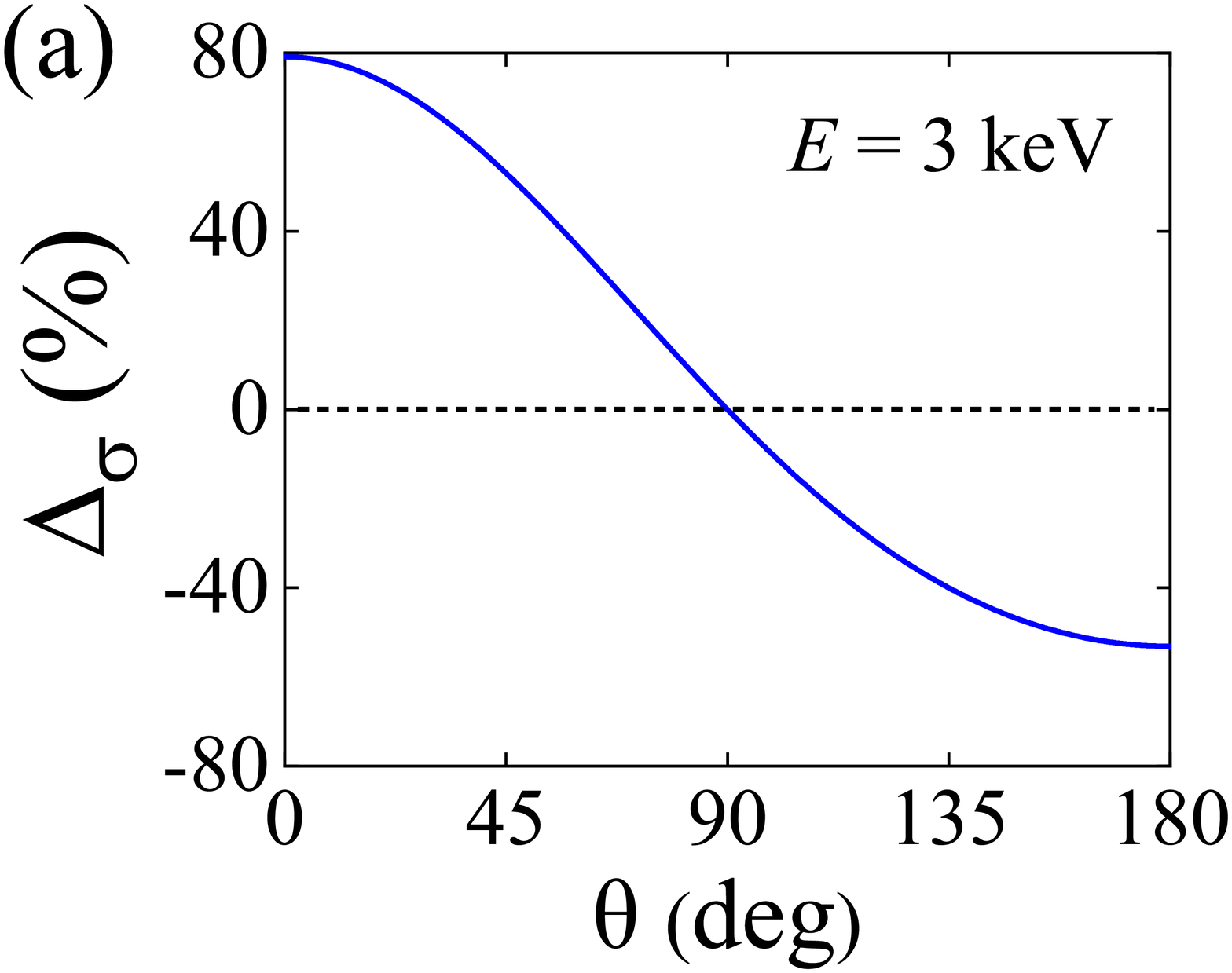}
	\end{minipage}
	\begin{minipage}[b]{0.23\textwidth}
		\includegraphics[width=1\textwidth]{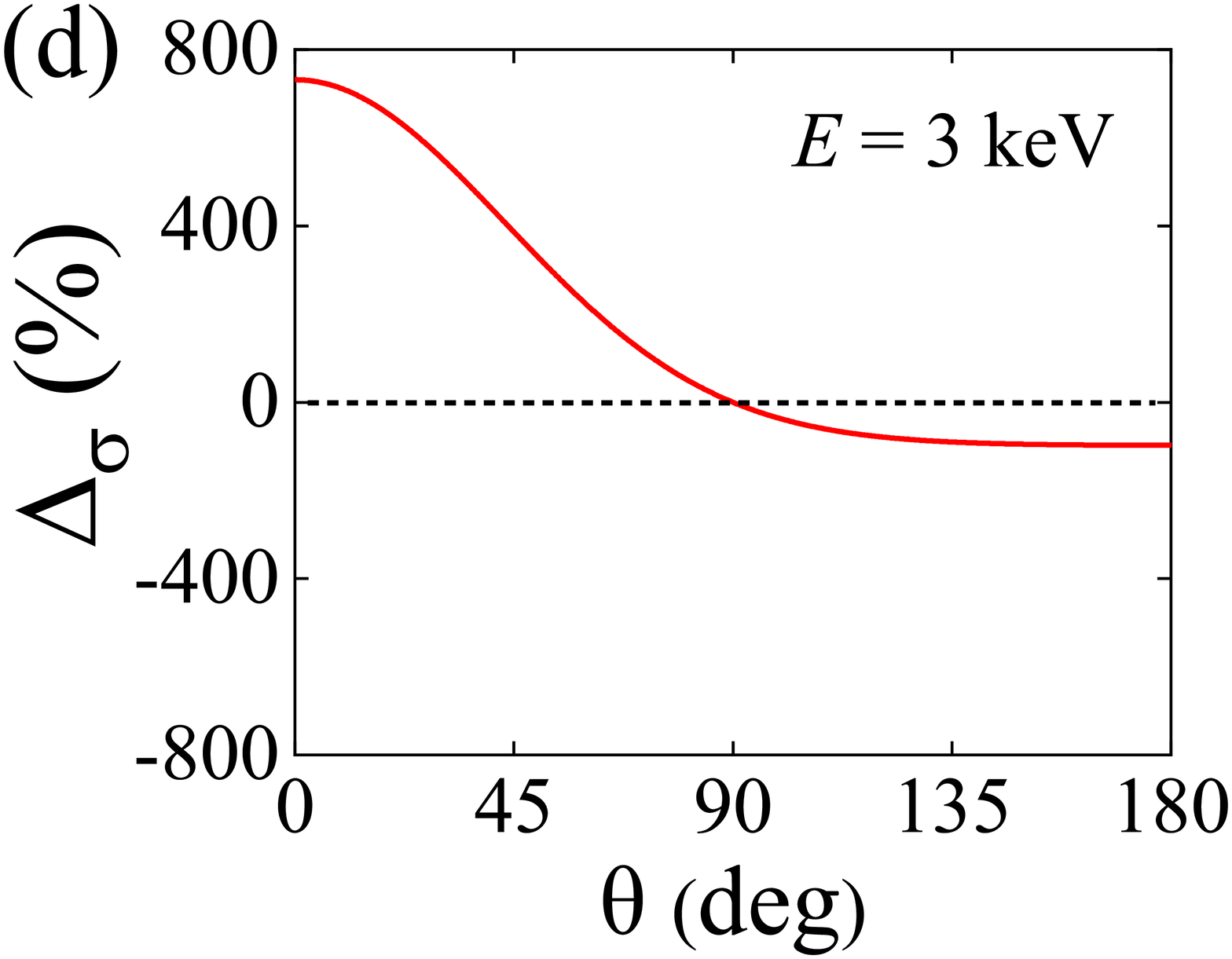}
	\end{minipage}
	\begin{minipage}[b]{0.23\textwidth}
		\includegraphics[width=1\textwidth]{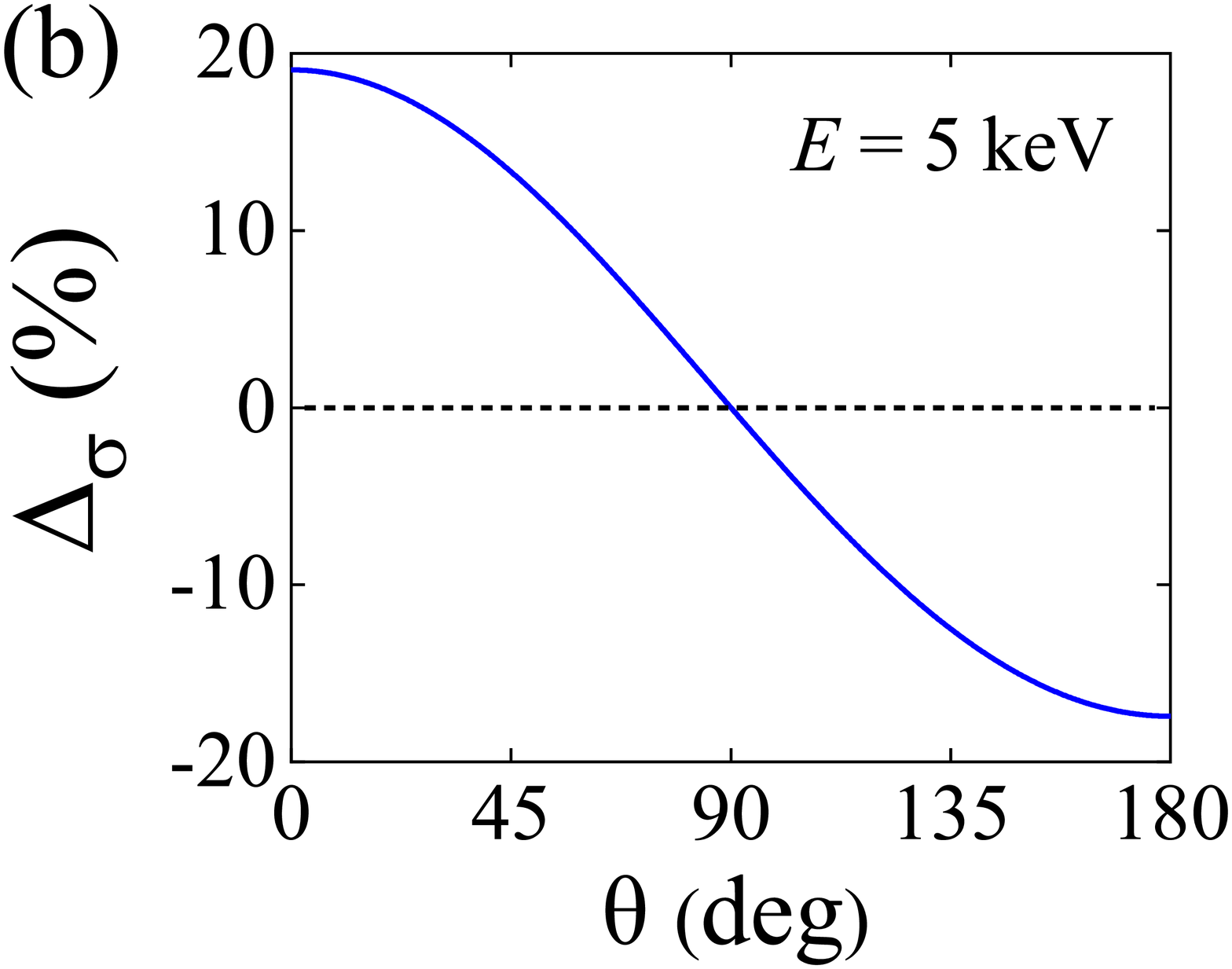}
	\end{minipage}
	\begin{minipage}[b]{0.23\textwidth}
		\includegraphics[width=1\textwidth]{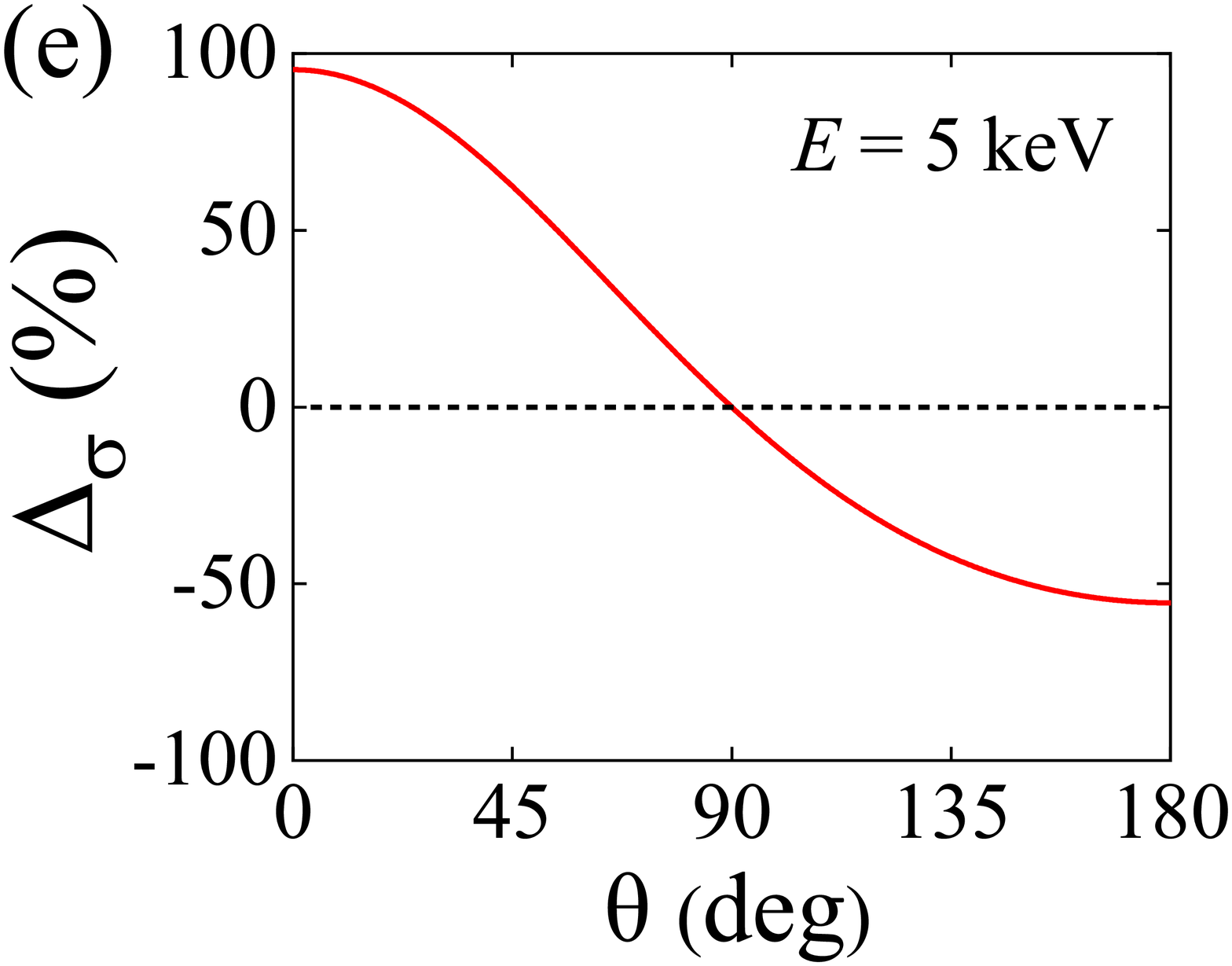}
	\end{minipage}
	\begin{minipage}[b]{0.23\textwidth}
		\includegraphics[width=1\textwidth]{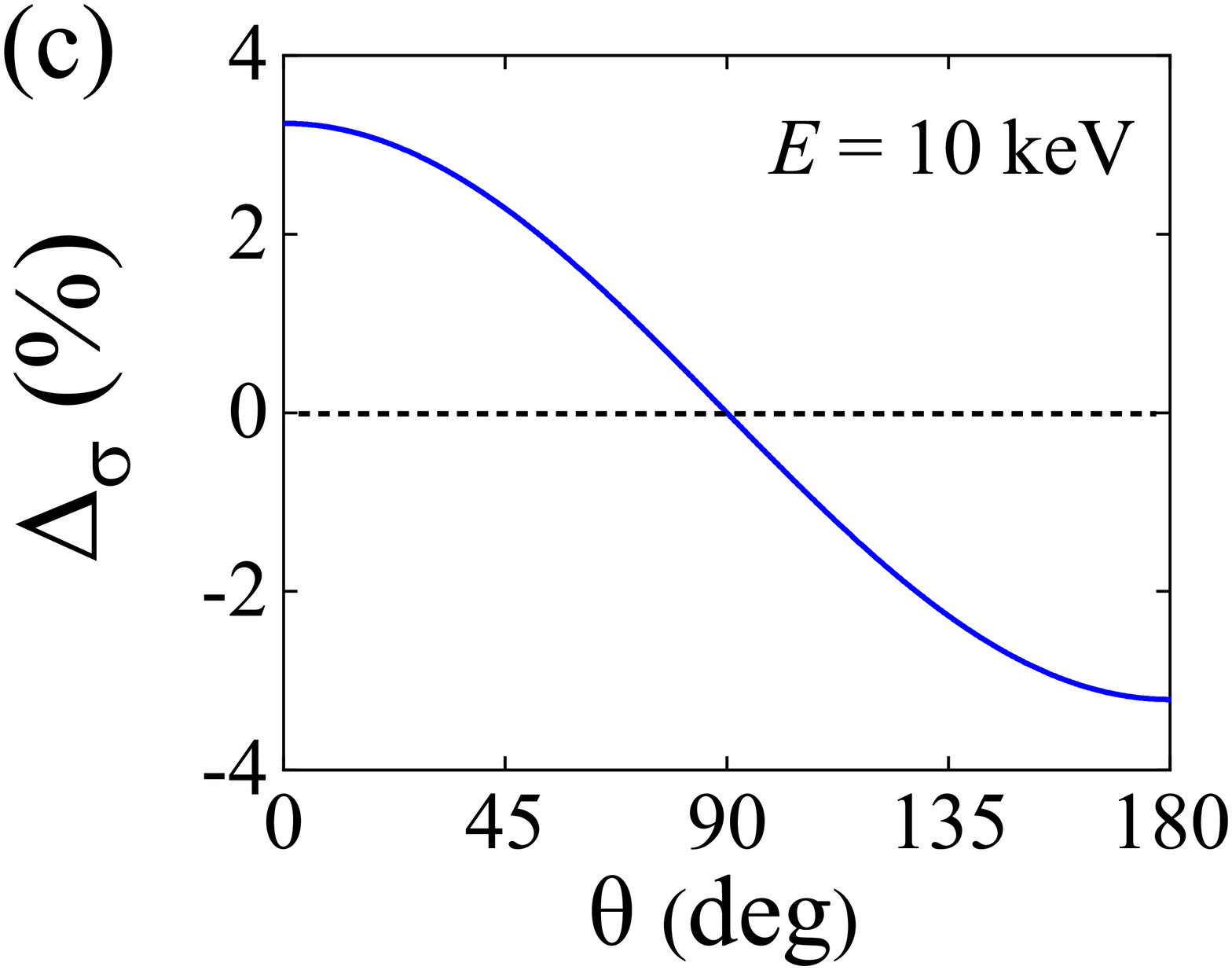}
	\end{minipage}
	\begin{minipage}[b]{0.23\textwidth}
		\includegraphics[width=1\textwidth]{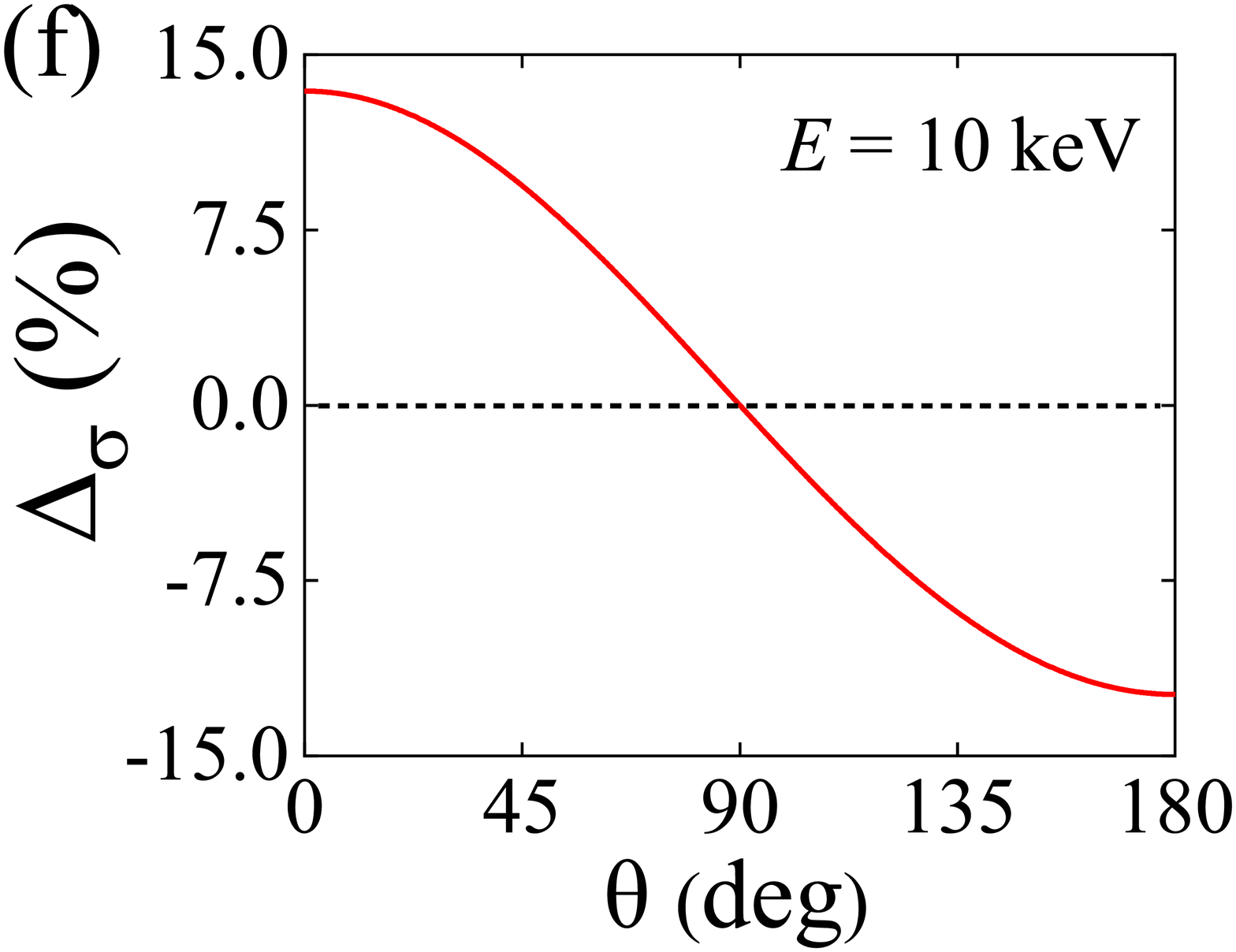}
	\end{minipage}

	\caption{The laser-induced modifications to the cross-sections for D-T fusion (blue line in panel (a), (b), and (c)) and D-$^{3}$He fusion (red line in panel (d), (e), and (f)), as a function of $\theta$. A peak intensity of 10$^{24}$ W/cm$^{2}$ is used.}
	\label{f.seta}
\end{figure}

It is shown that the cross-sections of thermonuclear fusion including D-T fusion and D-$^{3}$He fusion are very sensitive to the external intense laser fields at a peak intensity of 10$^{24}$ W/cm$^{2}$. Figure \ref{f.seta} shows the angular differential modifications to the cross-sections of D-T fusion (blue lines in panel (a), (b), and (c)) and D-$^{3}$He fusion (red lines in panel (d), (e), and (f)). The black dashed lines denote the relative change of the cross-section without laser fields. 

The positive-negative asymmetry between $\theta < 90^{\circ}$ and $\theta > 90^{\circ}$ can be seen in the panels of Figure \ref{f.seta}, especially at low energies. It is intuitive to find that the cross-section of D-$^{3}$He fusion is easier to be modified by intense laser fields. The positive-negative asymmetry for D-$^{3}$He fusion is more apparent than that for D-T fusion. The cross-sections of thermonuclear fusion can be strikingly modified at the direction of laser polarization. And there is no modification at the direction $\theta=90^{\circ}$ where no laser exists.

\subsection{Linear dependency on laser electric strength}

\begin{figure}[htbp]
	\centering
	\includegraphics[scale=0.3]{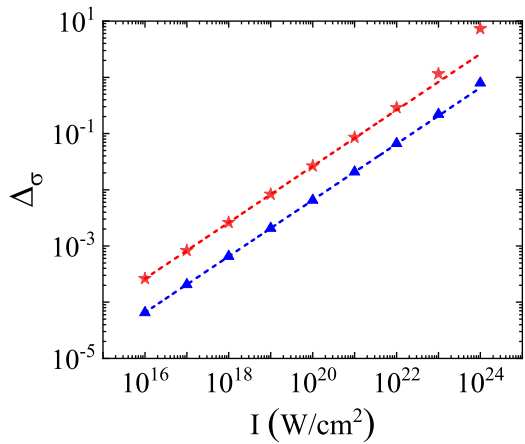}
	\caption{ The laser-induced modifications to the cross-sections for D-T fusion (blue triangles) and D-$^{3}$He fusion (red stars) at the energy $E = 3$ keV, as a function of laser intensity. }
	\label{f.intensity}
\end{figure}

Figure \ref{f.intensity} shows the laser-induced modifications to the cross-sections for D-T fusion (blue triangles) and D-$^{3}$He fusion (red stars) at the energy $E = 3$ keV, as a function of laser intensity. Both the laser-induced modifications $\Delta_{\sigma}$ and the laser intensity $I$ are plotted in the logarithmic scale. 

One notes the symbols are almost all dropped on the dashed lines with the slope of 0.5, due to a linear dependency on the laser electric strength. Here the relationship between laser electric strength and laser intensity is $\varepsilon$ [keV/fm] $=$ 2.744 $\times$ 10$^{-15}$ $\lbrace$I [W/cm$^{2}$]$\rbrace$$^{1/2}$.
With the laser intensities above 10$^{23}$ W/cm$^{2}$, the laser-induced modifications imply nonlinear dependency on laser electric strength which will be detailedly discussed latter in subsection D.

\subsection{Modifications within a typical range of temperatures}

\begin{figure}[htbp]
	\centering
	\includegraphics[scale=0.3]{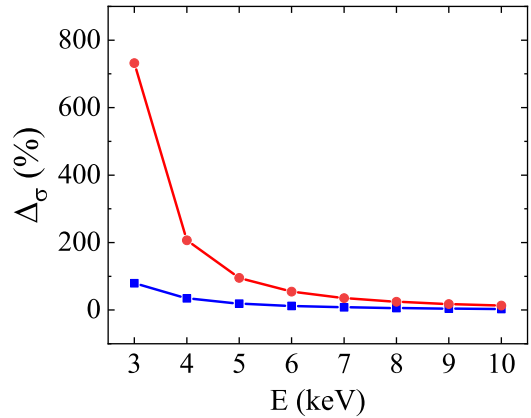}
	\caption{ The laser-induced modifications to the cross-sections for D-T fusion (blue squares) and D-$^{3}$He fusion (red circles), as a function of incident energy $E$. In this article, a range of temperatures of $3-10$ keV that people concerned in controlled fusion research is considered. A peak intensity of 10$^{24}$ W/cm$^{2}$ is used. }
	\label{f.energy}
\end{figure}

Figure \ref{f.energy} shows the laser-induced modifications to the cross-sections for D-T fusion (blue squares) and D-$^{3}$He fusion (red circles) when $\theta=0^{\circ}$, as a function of energy $E$. The laser-induced modification $\Delta_{\sigma}$ is plotted in the logarithmic scale. Results here are calculated within a range of energies $3-10$ keV which is the range of temperatures that people are concerned in controlled fusion research. And the considered laser intensity is 10$^{24}$ W/cm$^{2}$.

One sees the fusion cross-section at lower energy is easier to be modified by external laser fields. This is because the fusion reaction at lower energy has a longer tunneling path for the laser field to act on. 
At the energy $E=3$ keV, the laser-induced modifications to the cross-sections can reach about 79.1$\%$ for D-T fusion and 731.7$\%$ for D-$^{3}$He fusion. At the energy $E=10$ keV, the modifications to the cross-sections for D-T fusion and D-$^{3}$He fusion are 3.2$\%$ and 13.4$\%$, respectively.

\subsection{The generalized Gamow penetrability factor in intense laser fields}

It has been shown that how the external intense laser fields influence the thermonuclear fusion processes and how to calculate the laser-induced modification to the cross-section quantitatively. For the sake of understanding the results before, some analytical results will be given in this subsection. 

From the penetrability in intense laser fields of Eq. (\ref{eq.penetrability}), it is convenient to treat the laser interaction $V_{I}(r,\varepsilon,\theta)$ as a perturbation to the remaining potential $V_{0}(r)=V(r)-E$. And there is the Taylor expansion as 

\begin{align}
&P(E,\varepsilon,\theta)  \nonumber\\
&= \exp\left(-\frac{2\sqrt{2\mu}}{\hbar}\int_{R_{1}}^{R_{2}}\sqrt{V_{0}}\sqrt{1+\frac{V_{I}}{V_{0}}}dr\right),  \nonumber\\
&\approx\exp\left\lbrace-\frac{2\sqrt{2\mu}}{\hbar}\int_{R_{1}}^{R_{2}}\sqrt{V_{0}}\left(1+\frac{V_{I}}{2V_{0}}-\frac{V_{I}^{2}}{8V_{0}^{2}}\right)dr\right\rbrace, \nonumber\\
&= \exp( \gamma^{(0)}+\gamma^{(1)}+\gamma^{(2)} ),
\end{align}

where $\gamma^{(0)}$, $\gamma^{(1)}$, and $\gamma^{(2)}$ are defined as
\begin{align}
\gamma^{(0)}&=-\frac{2\sqrt{2\mu}}{\hbar}\int_{R_{1}}^{R_{2}}\sqrt{V_{0}}dr, \label{eq.g1}\\
\gamma^{(1)}&=\frac{\sqrt{2\mu}q_{eff}\varepsilon\cos\theta}{\hbar}\int_{R_{1}}^{R_{2}}\frac{rdr}{\sqrt{V_{0}}}, \label{eq.g2}\\
\gamma^{(2)}&=\frac{\sqrt{2\mu}(q_{eff}\varepsilon\cos\theta)^{2}}{4\hbar}\int_{R_{1}}^{R_{2}}\frac{r^{2}dr}{V_{0}^{3/2}}. \label{eq.g3}
\end{align}

One notes that the term $\gamma^{(0)}$ is independent on the laser field. But the terms $\gamma^{(1)}$ and $\gamma^{(2)}$ show a linear dependency and a quadratic dependency on the laser electric strength, respectively.

With the relationship $E=q_{1}q_{2}/R_{2}$, analytical solutions for Eq. (\ref{eq.g1}), Eq. (\ref{eq.g2}), and Eq. (\ref{eq.g3}) can be obtained by some mathematics.

First, for the term $\gamma^{(0)}$
\begin{align}
\gamma^{(0)}
&=-\frac{2\sqrt{2\mu}}{\hbar} E^{1/2} \int_{R_{1}}^{R_{2}}\sqrt{R_{2}/r-1}dr, \nonumber\\
&=-\frac{2\sqrt{2\mu}q_{1}q_{2}}{\hbar} E^{-1/2} \times \left[ \eta-\frac{1}{2}\sin (2\eta)\right], \nonumber\\
&\approx -B_{G} E^{-1/2},
\end{align}
where $\eta \equiv \cos^{-1}\sqrt{R_{2}/R_{1}} $, and a Puiseux serier expansion has been performed for the terms in the square bracket. 
The coefficient $B_{G}$ is given as 
\begin{align}
	B_{G} = \pi\sqrt{2\mu}q_{1}q_{2} / \hbar .
\end{align}
This is the so-called Gamow constant, and the term $\exp(\gamma^{(0)}$) is the Gamow penetrability factor without laser fields act on.

Second, for the term $\gamma^{(1)}$
\begin{align}
\gamma^{(1)}
&=\frac{\sqrt{2\mu}q_{eff}\varepsilon\cos\theta}{\hbar} E^{-1/2} \int_{R_{1}}^{R_{2}}\frac{rdr}{\sqrt{R_{2}/r-1}}, \nonumber\\
&=\frac{2\sqrt{2\mu}(q_{1}q_{2})^{2}q_{eff}\varepsilon\cos\theta}{\hbar}E^{-5/2} \nonumber\\
&\quad\times\left[ \frac{3}{8}\eta+\frac{1}{4}\sin(2\eta)+\frac{1}{32}\sin(4\eta)\right], \nonumber\\
&\approx C_{G}E^{-5/2}.
\end{align}
A Puiseux serier expsion has also been performed for the terms in the square bracket and the coefficient $C_{G}$ is 
\begin{align}
	C_{G} = 3\pi\sqrt{2\mu}(q_{1}q_{2})^{2}q_{eff}\varepsilon\cos\theta/(8\hbar).
\end{align}

\begin{table}[t]
	\centering
	\begin{spacing}{1.5}
		\caption{Laser-induced modifications to cross section. The $\Delta_{\sigma}$ values here are the same as those shown in Figure \ref{f.energy}. And the $\Delta_{\sigma_{G}}$ values are obtained using Eq. (\ref{eq.Gamow_delta}). The laser field is assumed to be at an intensity of 10$^{24}$ W/cm$^{2}$. } 
		\label{t.cross-section}
		\begin{tabularx}{8cm}{ p{1cm}<{\centering} | X<{\centering} | X<{\centering} | X<{\centering} | X<{\centering} }
			
			\hline\hline
			\bb{E\\(keV)} & \multicolumn{2}{c|}{ T(D, n)$\alpha$ } 			 & \multicolumn{2}{c}{ $^{3}$He(D, p)$\alpha$ } 		    \\ 
			\cline{2-5}
					&  $\Delta_{\sigma}$ $(\%)$  &	$\Delta_{\sigma_{G}}$ $(\%)$  &  $\Delta_{\sigma}$ $(\%)$   &  $\Delta_{\sigma_{G}}$ $(\%)$    \\
			\hline
			3 		&	 79.1	 				 & 76.9				 	 & 731.7				& 600.5						\\
			4 		&	 34.7					 & 34.4				 	 & 206.9				& 197.8						\\
			5 		&	 19.1					 & 19.0				  	 & 95.4					& 93.9					    \\
			6 		&	 11.9					 & 11.8				  	 & 54.6					& 54.2					    \\
			7 		&	 8.0					 & 8.0				 	 & 35.1					& 35.0					    \\
			8 		&	 5.7					 & 5.7				 	 & 24.3					& 24.3						\\
			9 		&	 4.2					 & 4.2				 	 & 17.7					& 17.7						\\
			10 		&	 3.2					 & 3.2				 	 & 13.4					& 13.4					    \\

			\hline
		\end{tabularx}
	\end{spacing}
\end{table}

Third, for the term $\gamma^{(2)}$
\begin{align}
\gamma^{(2)}
&= \frac{\sqrt{2\mu} (q_{eff} \varepsilon \cos\theta)^{2} }{2\hbar} \frac{\partial}{\partial E} \left(E^{-1/2}\int_{R_{1}}^{R_{2}}\frac{r^{2}dr}{\sqrt{R_{2}/r-1}}\right), \nonumber\\
&= -\frac{7\sqrt{2\mu} (q_{1}q_{2})^{3} (q_{eff} \varepsilon \cos\theta)^{2} }{2\hbar} E^{-9/2} \nonumber\\
&\quad\times\left[ \frac{5}{16}\eta+\frac{15}{64}\sin(2\eta)+\frac{3}{64}\sin(4\eta)+\frac{1}{192}\sin(6\eta)\right], \nonumber\\
&\approx -D_{G}E^{-9/2}.
\end{align}
Again, a Puiseux serier expsion has been performed from the second step to the third step. The coefficient $D_{G}$ is given as 
\begin{align}
	D_{G} = 35\pi\sqrt{2\mu} (q_{1}q_{2})^{3} (q_{eff}\varepsilon\cos\theta)^{2} / (64\hbar).
\end{align}

Finally, a generalized form of the Gamow penetrability factor in an intense laser field $\varepsilon$ can be obtained in the form as
\begin{equation} \label{eq.Gamow_penetrability}
P_{G}(E,\varepsilon,\theta) = \exp( -B_{G} E^{-1/2} +C_{G}E^{-5/2} -D_{G}E^{-9/2} ).
\end{equation}

The fusion cross-section based on the generalized Gamow penetrability factor can be written as 
\begin{equation} \label{eq.Gamow_cross-section}
	\sigma_{G}(E,\varepsilon,\theta)=\frac{S(E)}{E}P_{G}(E,\varepsilon,\theta).
\end{equation}
And the relative change of the cross-section is 
\begin{equation} \label{eq.Gamow_delta}
	\Delta_{\sigma_{G}} = \exp( C_{G}E^{-5/2} -D_{G}E^{-9/2} ) - 1.
\end{equation}

TABLE \ref{t.cross-section} shows a fairly high accuracy of the analytical results compared with the numerical results. Most $\Delta_{\sigma_{G}}$ values essentially agree with the $\Delta_{\sigma}$ values which are calculated using Eq. (\ref{eq.delta}). 

Results before in subsection A, B, and C then can be explained with analytical formulas. The positive-negative asymmetry of the angular differential modifications to the cross-section is due to the quadratic dependency on the laser field strength from the term $-D_{G}E^{-9/2}$. And the linear dependency on the laser field strength of the laser-induced modifications results from the term $C_{G}E^{-5/2}$. Especially at low energies and with extreme intensity exceeding 10$^{23}$ W/cm$^{2}$, the quadratic dependency on the laser field strength of modifications is quite remarkable. The generalized Gamow penetrability factor can be used to make more predictions on the laser-induced effects for some other similar nuclear processes.

\section{conclusions}

A theoretical method to study the possible influence of intense laser fields on thermonuclear fusion processes is shown in this article. It has been explained how the external laser fields influence the thermonuclear fusion processes and how to calculate the laser-induced modifications to cross-sections quantitatively. The quasistatic approximation is adopted in the study of thermonuclear fusion processes in the near-infrared intense lasers from the majority of intense laser facilities.

It is shown that the forthcoming intense laser fields can indeed make considerable modifications to the cross-sections for important thermonuclear reactions including D-T fusion and D-$^{3}$He fusion. 
The cross-section of D-$^{3}$He fusion is easier to be modified than that of D-T fusion.
At the temperature of 3 keV, the laser-induced modifications to the cross-sections can reach about 79.1$\%$ for D-T fusion and 731.7$\%$ for D-$^{3}$He fusion. And at the temperature of 10 keV, the modifications to the cross-sections for D-T fusion and D-$^{3}$He fusion are 3.2$\%$ and 13.4$\%$, respectively.

A linear dependency on the laser electric strength of the modifications to the cross-sections is reported. 
It is also reported that at low energies, the cross-section is very sensitive to the external laser fields and the quadratic dependency on the laser field strength of modifications are quite remarkable.
Finally, we give a generalized Gamow penetrability factor, a neat and fairly accurate analytical formula of the penetrability in intense laser fields, which is very useful to make more predictions on the laser-induced effects for some other similar nuclear processes.

\begin{acknowledgments}
	The author acknowledges fruitful discussions with Professor X. Wang and Professor L. B. Fu. This work was support by Science Challenge Project of China No. TZ2018005 and NSAF No. U1930403.
\end{acknowledgments}

\end{document}